\documentclass[aps,pra,twocolumn]{revtex4-1}
\usepackage{amsfonts,amssymb,amsmath}            % for math symbols.
\usepackage{lmodern}
\usepackage[]{graphics,graphicx}            % for graphics figures.
\usepackage{amsthm}
\usepackage{enumerate,mathtools,bbold}
\usepackage{qcircuit}
    \usepackage{hyperref}

\newcommand{\ket}[1]{\left | #1 \right\rangle}
\newcommand{\bra}[1]{\left \langle #1 \right |}
\newcommand{\half}{\mbox{$\textstyle \frac{1}{2}$}}

\newcommand{\braket}[2]{\left\langle #1|#2\right\rangle}
\newcommand{\proj}[1]{\ket{#1}\bra{#1}}
\newcommand{\identity}{\mathbb{1}}

\begin{document}
\title{Co-Processors for Quantum Devices}
\author{Alastair Kay}
\affiliation{Department of Mathematics, Royal Holloway University of London, Egham, Surrey, TW20 0EX, UK}
\email{alastair.kay@rhul.ac.uk}
\date{\today}
\begin{abstract}
Quantum devices, from simple fixed-function tools to the ultimate goal of a universal quantum computer, will require high quality, frequent repetition of a small set of core operations, such as the preparation of entangled states. These tasks are perfectly suited to realisation by a co-processor or supplementary instruction set, as is common practice in modern CPUs. In this paper, we present two quintessentially quantum co-processor functions: production of a GHZ state, and implementation of optimal universal (asymmetric) quantum cloning. Both are based on the evolution of a fixed Hamiltonian. We introduce a new technique for deriving the parameters of these Hamiltonians based on the numerical integration of Toda-like flows.
\end{abstract}
\maketitle

\section{Introduction} 

The task of quantum state synthesis \cite{kay2017,kay2017a} lies at the heart of any potential quantum technology -- before a quantum protocol can be run, be it a Bell test \cite{clauser1969}, quantum key distribution \cite{ekert1991}, quantum cloning \cite{buzek1996,werner1998,kay2009a}, random number generation \cite{pironio2010} or quantum computation \cite{raussendorf2001}, a non-trivial quantum resource must be prepared. The required resource might be a fixed quantum state such as a Bell state, $W$-state, cluster \cite{raussendorf2001} or GHZ state, or it might depend on a small input, such as the unknown state of a qubit. The availability of these resource states is the source of the power of quantum technologies. Repeated demands for the same resource state make it vital to concentrate on their accurate functioning. This suggests developing a device that accomplishes that single task, replacing a complex sequence of quantum gates. These might provide the first step in a quantum protocol (i.e.\ the core functionality of a particular quantum technological device), or operate as a fixed-function subroutine within a quantum computer, much as today's classical processors provide enhanced instruction sets (e.g.\ SSE or AVX) or co-processors. Our aim is to produce the desired states and transformations by the free evolution of a Hamiltonian whose parameters have been specifically tuned for the task. By doing this directly with the system's Hamiltonian for any relevant experimental scenario, whether this is in the solid state \cite{majer2007,plantenberg2007}, trapped ions \cite{islam2011}, or even photonic systems \cite{perez-leija2013,grafe2014,chapman2016}, we ensure that the state is produced as accurately and as quickly as possible, reducing the opportunities for external influences to degrade the resource.

  One special case of this has been extensively studied -- perfect quantum state transfer \cite{bose2003,christandl2004,christandl2005,kay2010a,difranco2008}, wherein an unknown quantum state can be transported between the two extremes of a one-dimensional chain of spins. This example demonstrates the power of the approach -- it is twice as fast as the equivalent quantum gate sequence \cite{yung2006} and many of the error modes are relegated to the manufacturing process; they can be identified prior to use and corrected, or simply rejected until a higher quality version is produced \cite{chapman2016}. In addition to perfect state transfer, the same device can create graph states \cite{clark2005}, of which the cluster states and GHZ states are special cases (albeit in an unusual basis). Minor modifications \cite{dai2010,kay2010a,genest2016} have also permitted the creation of Bell states between the extremal sites of the chain. More recently \cite{kay2017,kay2017a}, new systems have been created to facilitate the synthesis of arbitrary one-excitation states on the chain, such as $W$-states between subsets of sites, while more exotic interactions have been shown to cause signal amplification, ideal for enhancing measurement signals \cite{kay2007}.

In this paper, we develop a new co-processor that creates GHZ states in a particularly straightforward manner, see Section \ref{sec:GHZIsing}. Moreover, the one-dimensional transverse Ising model that we introduce is highly appropriate to many experimental scenarios from superconducting qubits \cite{majer2007,plantenberg2007} to trapped ions \cite{islam2011}, and demonstrates a reasonable robustness to experimental imperfections (see Section \ref{sec:robust}). In Section \ref{sec:cloning}, we also show how the state synthesis solutions of \cite{kay2017,kay2017a} can be combined with the GHZ co-processor to implement optimal universal cloning of one unknown qubit to $N$ clones \cite{kay2009a,kay2013,kay2016a}. This is the first time that a reliable implementation of optimal universal asymmetric cloning has been proposed (the circuits in \cite{kay2016a} were probabilistic in nature), demonstrating that fixed function devices can perform transformations based on a small input space, and realise highly non-trivial quantum properties. In Sec.\ \ref{sec:hypercube}, we also introduce the state synthesis problem for uniformly coupled networks (as compared to chains with engineered couplings), and demonstrate that some hypercubes are useful for generating the uniform superpositions ($W$-states) that are desirable for symmetric cloning.

Crucial to the specification of both co-processors is the numerical discovery of appropriate Hamiltonian parameters. We introduce a new technique based on the numerical integration of a differential equation, the Toda flow. This is the main focus of \ref{sec:synthesis}, also including discussions of convergence issues in Appendix \ref{sec:appendix}. Variants of this \cite{chu1986}, and good techniques for its numerical integration \cite{munthe-kaas1999,celledoni2014,iserles2000}, have been extensively studied in the numerical analysis and numerical methods literature.

\subsection{Perfect Excitation Transfer}

Throughout this work, we will rely on many of the insights previously developed in the study perfect state transfer. In essence, the core of this is that there is an $N\times N$ tridiagonal matrix
$$
h_{PST}^{(N)}=\left(\begin{array}{cccccc}
0 & J_1^{(N)} &&&&	\\
J_1^{(N)} & 0 & J_2^{(N)}&&&	\\
& J_2^{(N)} & 0 & J_3^{(N)} &&	\\
&&\ddots & \ddots & \ddots & \\
&&&J_{N-2}^{(N)}&0&J_{N-1}^{(N)}\\
&&&&J_{N-1}^{(N)}&0
\end{array}\right).
$$
This matrix has a basis $\{\ket{n}\}_{n=1}^N$, and the coupling strengths are chosen such that
\begin{equation}\label{eqn:pst}
e^{-ih_{PST}^{(N)}t_0}\ket{n}=(-i)^{N-1}\ket{N+1-n}
\end{equation}
The standard solution \cite{christandl2004} for these couplings is $J^{(N)}_n=\sqrt{n(N-n)}$ and $t_0=\pi/2$. Although there are infinitely many others, some analytic \cite{albanese2004}, some numeric \cite{karbach2005,kay2010a}, this first solution optimises many desirable features such as speed of transport \cite{yung2006,kay2006b,kay2016b}. On the other hand, \cite{karbach2005} provides an insightful method for producing coupling schemes that are close to some desirable configuration, perhaps imposed by experimental restrictions.

The key properties of $h_{PST}^{(N)}$ are related to its symmetry and its spectrum \cite{kay2010a}. The evolution time has to be long enough such that a relative phase of $\pi$ (modulo $2\pi$) is generated between neighbouring eigenvectors. As such, the minimum transfer time is related to the minimum eigenvalue gap $\Delta$ via $t_0\geq\pi/\Delta$. 

\section{GHZ State Creation}\label{sec:GHZIsing}

Systems such as the transverse Ising are now routinely accessed or simulated in quantum devices \cite{bernien2017}. The $N$-qubit Hamiltonian is
$$
H_I=\sum_{n=1}^NJ_nX_n+\sum_{n=1}^{N-1}B_nZ_nZ_{n+1},
$$
where $X_n$ and $Z_n$ denote the Pauli $X$ and $Z$ matrices respectively acting on qubit $n$.
We will now show how the parameters $J_n$ and $B_n$ can be tuned so that an initial separable state of $\ket{0}^{\otimes N}$, which is easily prepared, evolves to a maximally entangled GHZ state in fixed time. 

The evolution under $H_I$ is solved via a Jordan-Wigner transformation \cite{nielsen2005}. We invoke the Majorana fermions
$$
c_{2n-1}=X_1X_2\ldots X_{n-1}Z_n\qquad c_{2n}=X_1X_2\ldots X_{n-1}Y_n.
$$
These evolve under $H_I$ independently according to 
$$
c_n(t)=e^{-iH_It}c_ne^{iH_It}=\sum_{m=1}^{2N}c_m\bra{m}e^{-2ih_1t}\ket{n},
$$ where $h_1$ is the $2N\times 2N$ matrix
$$
h_1=i\left(\begin{array}{ccccccc}
0 & B_1 &&&&&	\\
-B_1 & 0 & J_1&&&&	\\
& -J_1 & 0 & B_2 &&&	\\
&&\ddots & \ddots & \ddots && \\
&&&-B_{N-1} & 0 & J_{N-1} & 0 \\
&&&&-J_{N-1}&0&B_N\\
&&&&&-B_N&0
\end{array}\right).
$$
Since the Majorana fermions form a basis, specifying the evolution invoked by this matrix $h_1$ fixes the evolution of the entire system. Moreover, the tridiagonal structure of $h_1$ is easily transformed into the form of a real symmetric tridiagonal which is well-studied for perfect state transfer \cite{bose2003,christandl2004,kay2010a}.

Recall that $J^{(N)}_n$ are the coupling strengths for a perfect state transfer scheme that has a transfer time $t_0$. By making the same identification as \cite{difranco2008},
$$
J_n=J^{(2N)}_{2n}\qquad B_n=J^{(2N)}_{2n-1},
$$
then $h_{PST}^{(2N)}$ satisfies Eq.\ (\ref{eqn:pst}), and 
$$
h_{PST}^{(2N)}=\left(\begin{array}{ccccccc}
0 & B_1 &&&&&	\\
B_1 & 0 & J_1&&&&	\\
& J_1 & 0 & B_2 &&&	\\
&&\ddots & \ddots & \ddots && \\
&&&B_{N-1} & 0 & J_{N-1} & 0 \\
&&&&J_{N-1}&0&B_N\\
&&&&&B_N&0
\end{array}\right)
$$
Introducing the unitary
$$
D=\sum_{n=1}^{2N}i^{n-1}\proj{n},
$$
we can transform between $h_{PST}^{(2N)}$ and $h_1$,
$$h_{PST}^{(2N)}=Dh_1D^\dagger.$$
This also updates the evolution,
\begin{multline*}
e^{-ih_1t_0}\ket{n}=D^{\dagger}e^{-ih_{PST}^{(2N)}t_0}D\ket{n}\\=(-i)^{2N-1}i^{n-1}D^{\dagger}\ket{2N+1-n}=(-1)^n\ket{2N+1-n}.
\end{multline*}
It follows that $c_n(t_0/2)=(-1)^nc_{2N+1-n}$.

We can now decompose the initial separable state in terms of the Majorana fermions, 
\begin{eqnarray}
\proj{0}^{\otimes N}&=&\frac{1}{2^N}\left(\prod_{n=1}^{N-1}(\identity+Z_nZ_{n+1})\right)(\identity+Z_1)	\nonumber\\
&=&\frac{1}{2^N}\left(\prod_{n=1}^{N-1}(\identity+ic_{2n}c_{2n+1})\right)(\identity+c_1). \label{eqn:projector}
\end{eqnarray}
After evolution under $H_I$ for time $t_0/2$, terms such as $c_{2n}c_{2n+1}$ transform into $-c_{2N+1-2n}c_{2N-2n}=c_{2N-2n}c_{2N+1-2n}$, via the anti-commutation of the fermions. Hence, the product involving pairs of fermions is unchanged, and the final state must become
$$
\frac{1}{2^N}\left(\prod_{n=1}^{N-1}(\identity+ic_{2n}c_{2n+1})\right)(\identity-c_{2N}).
$$
This is the same (up to a global phase) as the pure state
\begin{equation}
\ket{GHZ}=\frac{1}{\sqrt{2}}\left(\ket{0}^{\otimes N}-i\ket{1}^{\otimes N}\right).\label{eqn:ghz}
\end{equation}
We have successfully engineered an Ising chain that creates GHZ states by its natural dynamics. The transfer time scales linearly with $N$ if a maximum coupling strength is imposed, which is the best possible scaling for a one-dimensional system \cite{bravyi2006}.

The evolution after subsequent periods of $t_0/2$ is readily determined, since
$$
(\identity+c_1)\rightarrow(\identity-c_{2N})\rightarrow(\identity-c_1)\rightarrow(\identity+c_{2N})\rightarrow (\identity+c_1),
$$
so the state evolves as
\begin{equation}\label{eqn:sequence}
\ket{0}^{\otimes N}\rightarrow\ket{GHZ}\rightarrow-i\ket{1}^{\otimes N}\rightarrow -iZ_1\ket{GHZ}\rightarrow -i\ket{0}^{\otimes N}
\end{equation}
(indeed, the phase $Z_1$ might equally well be applied on any qubit). The only part that is not justified is the global phase. However, we know that
$$
\ket{0}^{\otimes N}\rightarrow\ket{GHZ}\qquad \ket{1}^{\otimes N}\rightarrow Z_1\ket{GHZ},
$$
so by linearity, we can evaluate $\ket{GHZ}\rightarrow-i\ket{1}^{\otimes N}$.

It is also now straightforward to determine the evolution of other basis states, which we will make use of in Sec.\ \ref{sec:cloning}. Consider an input state $\ket{x}$ for $x\in\{0,1\}^N$. We can alternatively write this as $X_x\ket{0}^{\otimes{N}}$ where $X_x$ has Pauli $X$ operators applied on the sites where the bit value of the string $x$ is 1, and identity on the other sites. Each $X_n=ic_{2n-1}c_{2n}$, and therefore evolves to $X_{N+1-n}$. So, if $x_R$ is the reversal of bit string $x$, we have
\begin{equation}
\ket{x}=X_x\ket{0}^{\otimes N}\mapsto X_{x_R}\ket{GHZ}.	\label{eqn:GHZmap}
\end{equation}

These models are well-suited to near-term experimental realisation. For example, \cite{bernien2017} uses Rydberg atoms in a chain to produce a Hamiltonian which can, in principle, be tuned to give an Ising model of up to 51 qubits. The main challenge is to make the fields $B_n$ and $J_n$ a similar strength. Currently, $|B_n/J_n|\leq\delta\lesssim 0.1$. While incompatible with the standard perfect state transfer couplings \cite{christandl2004}, other techniques such as \cite{karbach2005} can return the couplings for suitable perfect transfer schemes. However, this means that there are two eigenvalues of $h_1$ which are separated by $O(\delta^N)$ (corresponding to the unpaired fermions of the Majorana chain described by Kitaev \cite{kitaev2001}). Consequently, the time for generating the GHZ state scales as $\Omega(\delta^{-N})$, which is currently prohibitive. %While there are long-range couplings in \cite{bernien2017}, since their strengths drop with $r^{-6}$ between spins separated by distance $r$, we can neglect terms beyond nearest-neighbour. For example, a chain of 11 spins still produces a state with overlap of approximately 97\% with the target GHZ state.

\subsection{Robustness of GHZ Synthesis}\label{sec:robust}

\begin{figure}
\begin{center}
\includegraphics[width=0.45\textwidth]{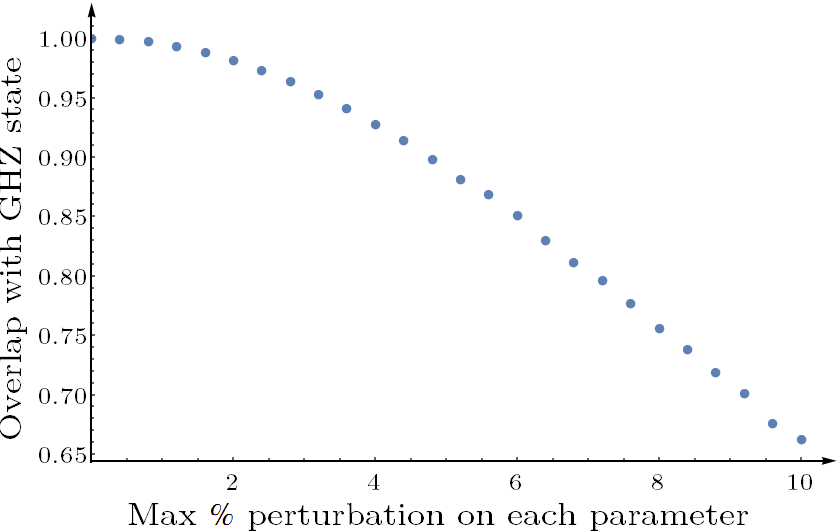}
\caption{Average overlap with $\ket{GHZ}$ for $N=21$ when the parameters $B_n$ and $J_n$ are all chosen uniformly at random within a range of $\pm x\%$ of the value they should be (the `standard' perfect transfer couplings, \cite{christandl2004}), using 1000 samples.}\label{fig:perturbed}
\vspace{-0.5cm}
\end{center}
\end{figure}

While we have identified the main experimental challenge, the necessary accuracy of engineering for the system parameters could be a concern. This is particularly acute in the case of GHZ state synthesis, because the state in Eq.\ (\ref{eqn:projector}) is described in terms of a large number (up to $2N-1$) of Majorana fermions. If each is transported with some sub-unital fidelity, the overall success of the synthesis could be quite minimal. In fact, the situation is not nearly so bleak, and the system has a good tolerance of these imperfections. A good estimate on the overlap of the output state $\bra{GHZ}e^{-iH_It_0/2}\ket{0}^{\otimes N}$ is
$$
\approx\frac{1+|F|}{2^N}\sqrt{\text{det}\left(e^{-ih_1t_0}h_0e^{ih_1t_0}h_0-\identity\right)},
$$
where $F=\bra{2N}e^{-ih_1t_0}\ket{1}$ is the single excitation transfer fidelity, and
$$
h_0=\sum_{n=1}^{N-1}\ket{2n}\bra{2n+1}-\ket{2n+1}\bra{2n}.
$$
This exactly evaluates the evolution of $\prod_{n=1}^{N-1}(\identity+Z_nZ_{n+1})$, comparing it to itself, by writing it as a fermionic Gaussian state \cite{bravyi2005,kay2011}. The additional effect of the single excitation $c_1$ is then approximated, ignoring possible interactions with the Gaussian component. This approximation facilitates numeric simulations, and Fig.\ \ref{fig:perturbed} demonstrates the effect on a chain of size 21.

It must be emphasised that these results are very basic, merely measuring the overlap with the target state. When this device is made, we will characterise the state that is produced, and adapt for its imperfections, such as applying an optimised choice of local unitaries. This can only serve to increase the figure of merit. Or, can we witness the presence of different types of entanglement? In particular, $k$-body entanglement for $k\sim N$. Existing entanglement witnesses are not yet sophisticated enough to be able to discriminate this.

On the other hand, one thing that we cannot easily do is replace the perfect transfer couplings with a coupling scheme that achieves nearly perfect transfer (but with a shorter transfer time, making the system less susceptible to noise, and some perturbations), such as those suggested in \cite{apollaro2012}. Those schemes are tuned specifically for end-to-end transfer. They generate high transfer fidelity between $\ket{1}\rightarrow\ket{2N}$ at a higher speed, at the cost of the transfer fidelity between intermediate sites. However, GHZ synthesis requires high quality transfer for all pairs $\ket{n}\rightarrow\ket{2N+1-n}$. For example, the optimal $N=21$ solution from \cite{apollaro2012} has a $1\rightarrow 42$ excitation transfer fidelity of 0.993 (which is roughly reproduced by the $3\%$ perturbed chains in Fig.\ \ref{fig:perturbed}), but only generates the GHZ state with overlap 0.762 due to the vastly lower transfer fidelities on the middle of the chain, such as for $5\rightarrow 38$, which is less than 0.4. 

\subsection{GHZ Creation in the XY Model}\label{secGHZXY}

A generalisation of $H_I$ can be written as:
$$
H_{ZY}\!\!=\!\!\sum_{n=1}^N\!B_nX_n+\!\sum_{n=1}^{N-1}\!J_{n}(1+\gamma_n)Z_nZ_{n+1}+J_n(1-\gamma_n)Y_nY_{n+1}.
$$
This model is also a free-fermion model, and has the same Majorana fermions as $H_I$. Starting from the same initial state, as described by Eq.\ (\ref{eqn:projector}), the only possible difference is what those Majorana fermions can evolve into, which is again governed by a $2N\times 2N$ matrix $h_\gamma$, similar to $h_1$. We are interested in whether this broader class of Hamiltonians can also produce the GHZ state, again in the hope of improving experimental viability. For pedagogical simplicity, we will fix $\gamma_n=\gamma$ for all $n$, although there is no such restriction arising in the mathematics. We could equally well consider the Hadamard-transformed version of this Hamiltonian, which is the more familiar $XY$ model.
$$
H_{XY}\!\!=\!\!\sum_{n=1}^N\!B_nZ_n+\!\sum_{n=1}^{N-1}\!J_{n}(1+\gamma_n)X_nX_{n+1}+J_n(1-\gamma_n)Y_nY_{n+1}.
$$
In this case, the initial state would be $(H\ket{0})^{\otimes N}$, and the final state would be $H^{\otimes N}\ket{GHZ}$.

We already know two solutions for this matrix. At $\gamma=1$ (vanishing $YY$ terms), we have already fixed $B_n=J^{(2N)}_{2n-1}$ and $J_n=J^{(2N)}_{2n}/2$, while for $\gamma=0$, $B_n=N$ and $J_n=J^{(N)}_n/2$ comes from perfect state transfer. Indeed, this last solution is the usual perfect state transfer chain (using $H_{XY}$), and it was already observed in \cite{clark2005} that this system is capable of creating the GHZ state (in a non-obvious basis).

Solutions for both values of $\gamma$ have the same eigenvalues, $\pm 1,\pm 3,\pm 5,\ldots \pm(2N-1)$. We are hence interested at intermediate values of $\gamma$, with the same spectrum. We shall do this by providing a numerical routine to interpolate between the two known solutions. We believe the form of the isospectral transformation is new to the spin chain community, although is well-studied in the numerical methods and analysis literature \cite{chu1986,munthe-kaas1999,celledoni2014,iserles2000}.

We permute the elements of $h_{\gamma}$, grouping odd-numbered and even-numbered basis elements together. The matrix then decomposes as
$$
h_\gamma=i\ket{0}\bra{1}\otimes X(\gamma)-i\ket{1}\bra{0}\otimes X^T(\gamma)
$$
where $X(\gamma)$ is a non-symmetric matrix
\begin{multline*}
X=\sum_{n=1}^NB_n\proj{n}+\sum_{n=1}^{N-1}J_n(1+\gamma)\ket{n}\bra{n+1}\\+\sum_{n=1}^{N-1}J_n(1-\gamma)\ket{n+1}\bra{n}.
\end{multline*}
The spectrum of $h_{\gamma}$ is $\pm\lambda_i$ where $\lambda_i$ are the singular values of $X$. Hence, it is sufficient to perform an isospectral transformation on $X$. To achieve this, we observe that if $A$ and $B$ are anti-Hermitian, then
$$
X(t)=e^{-B}Xe^{A}
$$ 
describes an isospectral flow. Taking the derivative,
\begin{equation} \label{eqn:flow2}
\frac{dX}{dt}=XA-BX.
\end{equation}
So, any small, anti-Hermitian $A$, $B$ will achieve an isospectral transformation, we just need to select them so that $X$ retains the properties that we want it to:
\begin{enumerate}
\item $X$ is tridiagonal, i.e.\ $\bra{m}XA\ket{n}=\bra{m}BX\ket{n}$ for any $n,m$ such that $|n-m|>1$.
\item $C$ is centrosymmetric, in the sense $\bra{n}X\ket{m}=\bra{N+1-m}X\ket{N+1-n}$. We anticipate this being a necessary condition in the same way that it is for state transfer \cite{kay2010a}, although this is unproven.
\item We require $\bra{n+1}X\ket{n}/\bra{n}X\ket{n+1}=\frac{1-\gamma}{1+\gamma}$ to be the same for all $n$.
\end{enumerate}
As conditions on the matrix $X$, if it satisfies them at any given value, we can ensure they are upheld on subsequent values by imposing them on the derivatives.
For a given $X$, each condition is linear in the coefficients of $A$ and $B$, and the number of constraints coincides perfectly with the number of coefficients, permitting solution. By performing a numerical integration starting from a known solution for $X(0)$ (which, as already observed, we know for $\gamma=0,1$) any desired value of $\gamma$ can be arrived at.

Equation (\ref{eqn:flow2}) can be integrated following two different philosophies. Firstly, one can integrate it directly, i.e.\ setting the next $X$ to be
$$
X\mapsto X+\delta(XA-BX).
$$
The structural aspects of $X$ are preserved exactly, but the isospectral transformation is only accurate to $O(\delta^2)$ for each step of size $\delta$, giving an overall accuracy of $O(\delta)$. Alternatively, one can perform the update $$X\rightarrow e^{-B}Xe^A.$$
This unitary transformation is isospectral, but the structural properties such as tridiagonality are only accurate to $O(\delta^2)$. Nevertheless, there is the facility to compensate for any error in the next step, preventing it from accumulating during the integration. Moreover, if $\delta$ is shrunk as a solution converges, then the accuracy is arbitrarily good. Throughout this paper, we use first-order (Euler) integrations. While they appear to serve very well, isospectral flows of this form are often challenging to integrate \cite{calvo1997}, and novel techniques such as Runge-Kutta-Munthe-Kaas have been developed \cite{munthe-kaas1999,celledoni2014}. These may be used to improve performance in the future.

We conclude that any model $H_{ZY}$ can be tuned to achieve GHZ state generation.
An explicit demonstration is given in \cite{kay2017b} for 21 qubits and $\gamma=0.7$ (integrating from $\gamma=0$, using the direct method). The modest choice of size derives only from memory limits of simulating a full Hamiltonian for verification.

\section{Optimal Cloning}\label{sec:cloning}

Production of the GHZ state has shown that although studies of state transfer are often constrained to the single excitation subspace, the same ideas can be applied to generate interesting evolution in multiple excitation sectors. For the GHZ state, this was a fixed input providing a fixed output. Are there other protocols that we might consider? An arbitrary unitary seems to be out of the question, even within the single excitation subspace -- if we use a state synthesis routine \cite{kay2017,kay2017a} then one can choose the evolution of a particular excitation, say $\ket{1}\rightarrow\ket{\psi}$. But then, the possible evolution of other input states is tightly constrained. For example, since $\ket{2}=H\ket{1}/J_1$,
$$
e^{-iHt}\ket{2}=\frac{1}{J_1}He^{-iHt}\ket{1}=\frac{1}{J_1}H\ket{\psi}.
$$
So, if $\ket{\psi}$ is confined to a small set of sites, $\ket{2}$ evolves only onto those and neighbouring sites. Perfect state transfer demonstrates this -- any system that transfers $\ket{1}\rightarrow\ket{N}$ must also transfer $\ket{n}\rightarrow\ket{N+1-n}$; there is no freedom to choose these transformations. Still, there may be interesting protocols that depend upon a small input subspace. Perfect state transfer is one such example, wherein the possible inputs are spanned by a basis of 2 states. Another example is $1\rightarrow N$ cloning. The optimal $1\rightarrow N$ universal  asymmetric cloning machine \cite{kay2009a,kay2013,kay2016a} implements the (not necessarily unique) transformation
\begin{equation}\label{eqn:clonemap}
\begin{aligned}
\ket{0}&\mapsto A\ket{0}^{\otimes N}+\sum_{n=1}^N\beta_n\ket{1}^{\otimes(n-1)}\ket{0}\ket{1}^{\otimes(N-n)}\\
\ket{1}&\mapsto A\ket{1}^{\otimes N}+\sum_{n=1}^N\beta_n\ket{0}^{\otimes(n-1)}\ket{1}\ket{0}^{\otimes(N-n)}	
\end{aligned}
\end{equation}
where
$$
A=\sum_{n=1}^N\beta_n\qquad B^2=\sum_{n=1}^N\beta_n^2,
$$
and the $\beta_n$ determine the asymmetry of the cloning quality via the single-copy average fidelities
$$
F_n=\frac{2+\left(\beta_n+A\right)^2}{6},
$$
and satisfy the normalisation
$
A^2+B^2=1.
$
We will now show how this can be implemented using spin chains. There is not a single spin chain that achieves this entire transformation, but we can use them as tools that massively simplify the sequence of quantum gates that need to be applied. To that end, consider a set of $M=2N-1$ qubits. One qubit, $k+2$, is the unknown state to be cloned, $\ket{\psi}$, but rotated by a phase gate, and we still aim to produce $N$ clones, on the odd-numbered qubits. The rest are prepared in the separable state:
$$
\ket{0}^{\otimes k}\left(A\ket{0}+iB\ket{1}\right)\otimes\sqrt{Z}\ket{\psi}\otimes\ket{0}^{\otimes(M-2-k)}.
$$
This can be decomposed into the four basis states
$$
\ket{0}^{\otimes k}\ket{x}\ket{0}^{\otimes(M-2-k)}
$$
for $x\in\{0,1\}^2$. We evolve with any $M$-qubit GHZ-generating $H_{ZY}$ for the generation time. According to Eq.\ (\ref{eqn:GHZmap}), this produces the states
$$
X_{M-k}^{x_1}X_{M-k-1}^{x_2}\ket{GHZ}.
$$
Now, we apply a controlled-phase gate between the qubits $M-k$ and $M-k-1$. The effect is that if the two bits of $x$ are the same value, this works like a phase gate on the $\ket{GHZ}$ state (skipping us on 2 steps in Eq.\ (\ref{eqn:sequence})), while it does nothing if the two bit values are different. Then, we repeat the GHZ evolution. Referring to Eq.\ (\ref{eqn:sequence}), this returns $X_{k+1}^{x_1}X_{k+2}^{x_2}\ket{0}^{\otimes M}$ if the two bits of $x$ are equal, and $X_{k+1}^{x_1}X_{k+2}^{x_2}\ket{1}^{\otimes M}$ otherwise. Overall,
\begin{eqnarray*}
\ket{0}&\mapsto&A\ket{0}^{\otimes M}+BX_{k+1}\ket{1}^{\otimes M}	\\
\ket{1}&\mapsto&AX_{k+2}\ket{1}^{\otimes M}+BX_{k+1}X_{k+2}\ket{0}^{\otimes M}
\end{eqnarray*}
has been implemented (the inputs being the basis of the qubit to be cloned). Finally, we apply a controlled-NOT controlled from qubit $k+1$ and targeting qubit $k+2$. This gives the overall transformation
\begin{eqnarray*}
\ket{0}&\mapsto&A\ket{0}^{\otimes M}+B\ket{1}^{\otimes k}\ket{0}\ket{1}^{\otimes(M-1-k)}	\\
\ket{1}&\mapsto&A\ket{1}^{\otimes M}+B\ket{0}^{\otimes k}\ket{1}\ket{0}^{\otimes (M-k-1)}
\end{eqnarray*}
In fact, the entire transformation up to this point can be implemented by a single Hamiltonian evolution, using a less physically motivated Hamiltonian, based on a tuned 3-body cluster state Hamiltonian \cite{kay2007}. Alternatively, as already observed, one can replace the Ising-generating Hamiltonian with a perfect state transfer Hamiltonian by applying a Hadamard transform before and after. This has the advantage of making the form of the Hamiltonian for the GHZ generation and state synthesis parts the same, up to modification of the coupling strengths, at the cost of adding some local Hadamard gates.

\begin{figure}[!tbp]
\begin{center}
\includegraphics[width=0.4\textwidth]{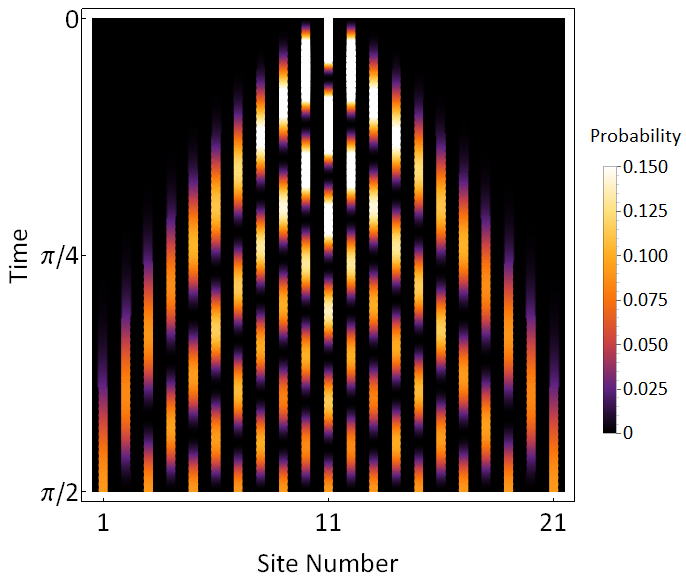}
\end{center}
\vspace{-0.5cm}
\caption{A single excitation, input to the central spin of a 21-qubit spin chain (top) evolves into a uniform superposition over the odd-numbered sites (bottom), as required for optimal symmetric universal $1\rightarrow 11$ cloning.}\label{fig:WstateOddRevival}
\vspace{-0.3cm}
\end{figure}

From here, we can get our overall desired cloning transformation, creating the $N$ clones on the odd-numbered qubits of the chain, if we can implement
\begin{eqnarray*}
\ket{0}^{\otimes M}&\mapsto&\ket{0}^{\otimes M}	\\
\ket{1}^{\otimes M}&\mapsto&\ket{1}^{\otimes M}	\\
\ket{k+1}&\mapsto&\frac{\sum_{n=1}^N\beta_n\ket{2n-1}}{B}	\\
\ket{\overline{k+1}}&\mapsto&\frac{\sum_{n=1}^N\beta_n\ket{\overline{2n-1}}}{B},
\end{eqnarray*}
where $\ket{0}^{\otimes (k-1)}\ket{1}\ket{0}^{\otimes (M-k)}=\ket{k}$ and $\ket{\bar k}=X^{\otimes M}\ket{k}$.
The first two transformations are automatic for an exchange-coupled spin chain
\begin{equation}
H_{XX}=\sum_{n=1}^{M-1}\frac{J_n}{2}(X_nX_{n+1}+Y_nY_{n+1})	\label{eqn:exchange}
\end{equation}
because $\ket{0}^{\otimes M}$ and $\ket{1}^{\otimes M}$ are null vectors of $H_{XX}$. Assume that couplings can be found in order to implement the third transformation in a time $t_0$ \cite{kay2017,kay2017a}. This will be discussed in Sec.\ \ref{sec:synthesis}. Indeed, a suitable solution was found in \cite{kay2017a} for an initial state in the middle, $k=N-1$, and is reproduced in Fig.\ \ref{fig:WstateOddRevival}. For the last condition, observe that $[H_{XX},X^{\otimes M}]=0$. Hence
$$
e^{-iH_{XX}t_0}\ket{\bar k}=X^{\otimes M}e^{-iH_{XX}t_0}\ket{k},
$$
which simply yields that the transformations in the 1 and $M-1$ excitation subspaces are essentially identical, and so the final condition will also be satisfied.

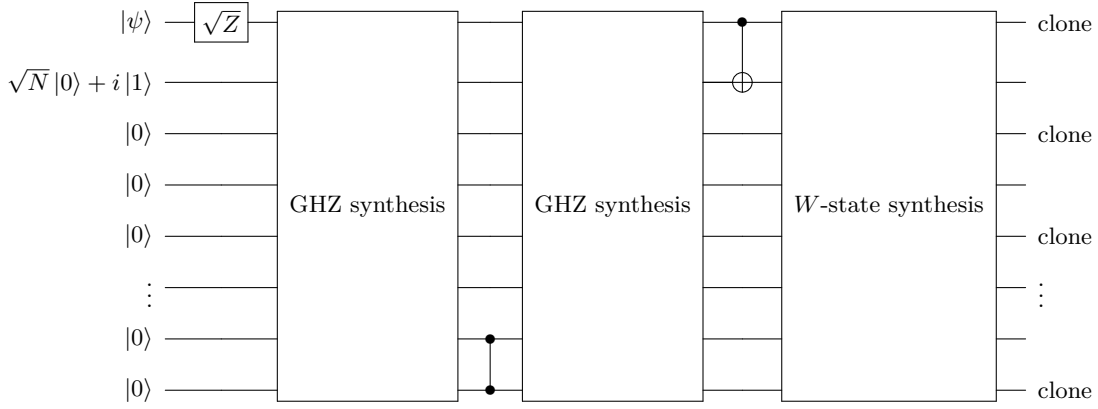
\begin{figure*}[!tbp]
$$
\Qcircuit @C=1.2em @R=1.2em{
\lstick{\ket{\psi}}& \gate{\sqrt{Z}}& \multigate{7}{\text{GHZ synthesis}}& \qw & \multigate{7}{\text{GHZ synthesis}} &\ctrl{1} & \multigate{7}{W\text{-state synthesis}} & \rstick{\text{clone}} \qw	\\
\lstick{\sqrt{N}\ket{0}+i\ket{1}}& \qw& \ghost{\text{GHZ synthesis}}& \qw  & \ghost{\text{GHZ synthesis}}& \targ \qw & \ghost{W\text{-state synthesis}} &  \qw	\\
\lstick{\ket{0}}& \qw& \ghost{\text{GHZ synthesis}}& \qw & \ghost{\text{GHZ synthesis}} &\qw & \ghost{W\text{-state synthesis}}	& \rstick{\text{clone}} \qw		\\
\lstick{\ket{0}}& \qw& \ghost{\text{GHZ synthesis}}& \qw & \ghost{\text{GHZ synthesis}} &\qw & \ghost{W\text{-state synthesis}} &  \qw				\\
\lstick{\ket{0}}& \qw& \ghost{\text{GHZ synthesis}}& \qw & \ghost{\text{GHZ synthesis}} &\qw & \ghost{W\text{-state synthesis}}	& \rstick{\text{clone}} \qw		\\
\lstick{\vdots}& \qw& \ghost{\text{GHZ synthesis}}& \qw & \ghost{\text{GHZ synthesis}} &\qw & \ghost{W\text{-state synthesis}} & \rstick{\vdots} \qw				\\
\lstick{\ket{0}}& \qw& \ghost{\text{GHZ synthesis}}&\ctrl{1} & \ghost{\text{GHZ synthesis}} &\qw & \ghost{W\text{-state synthesis}} &  \qw					\\
\lstick{\ket{0}}& \qw& \ghost{\text{GHZ synthesis}}&\control \qw& \ghost{\text{GHZ synthesis}} &\qw & \ghost{W\text{-state synthesis}}	& \rstick{\text{clone}} \qw			\\
}
$$
\caption{Quantum circuit diagram for quantum cloning when supplemented by two helper Hamiltonians, acting on $M=2N-1$ input qubits. Coefficients are specifically chosen for optimal universal symmetric cloning. The input qubit can be arbitrary.}\label{fig:QcircuitHam}

\end{figure*}

\begin{figure*}[!tbp]
$$
\Qcircuit @C=1.2em @R=1.2em{
\lstick{\ket{\psi}}& \ctrl{7} & \qw &\ctrl{1}	& \multigate{1}{\sqrt{\text{\sc{swap}}}} & \qw & \multigate{1}{\sqrt{\text{\sc{swap}}}} & \qw & \qw& \multigate{1}{\sqrt{\text{\sc{swap}}}} & \qw & \rstick{\text{clone}} \qw	\\
\lstick{\sqrt{N}\ket{0}+\ket{1}}& \qw & \ctrl{6} & \targ & \ghost{\sqrt{\text{\sc{swap}}}} & \qswap& \ghost{\sqrt{\text{\sc{swap}}}} & \qswap & \qw& \ghost{\sqrt{\text{\sc{swap}}}} & \gate{\sqrt{Z}^\dagger} & \rstick{\text{clone}} \qw	\\
\lstick{\ket{0}}& \targ & \targ& \qw  & \qw & \qswap \qwx & \multigate{1}{\sqrt{\text{\sc{swap}}}} & \qw & \qw& \multigate{1}{\sqrt{\text{\sc{swap}}}} & \gate{\sqrt{Z}^\dagger}	& \rstick{\text{clone}} \qw		\\
\lstick{\ket{0}}& \targ & \targ& \qw	 & \qw& \qw& \ghost{\sqrt{\text{\sc{swap}}}} & \qw & \qswap& \ghost{\sqrt{\text{\sc{swap}}}} &\gate{Z} & \rstick{\text{clone}} \qw				\\
\lstick{\ket{0}}& \targ & \targ& \qw  & \qw& \qw& \qw & \qswap \qwx[-3] & \qw& \multigate{1}{\sqrt{\text{\sc{swap}}}} & \gate{\sqrt{Z}^\dagger}	& \rstick{\text{clone}} \qw		\\
\lstick{\ket{0}}& \targ & \targ& \qw	 & \qw& \qw& \qw & \qw & \qw& \ghost{\sqrt{\text{\sc{swap}}}} & \gate{Z} & \rstick{\text{clone}} \qw				\\
\lstick{\vdots}& \targ & \targ& \qw & \qw& \qw& \qw & \qw & \qswap \qwx[-3]& \multigate{1}{\sqrt{\text{\sc{swap}}}} & \gate{Z} & \rstick{\vdots} \qw					\\
\lstick{\ket{0}}& \targ & \targ& \qw	 & \qw& \qw& \qw & \qw & \qw& \ghost{\sqrt{\text{\sc{swap}}}} & \gate{\sqrt{Z}}	& \rstick{\text{clone}} \qw			\\
}
$$
\caption{Quantum circuit diagram for quantum cloning without the two helper Hamiltonians. Moving beyond the symmetric cloner requires replacement of the $\sqrt{\text{\sc{swap}}}$ with partial {\sc swap} operations of different amounts.}\label{fig:QcircuitGates}

\end{figure*}
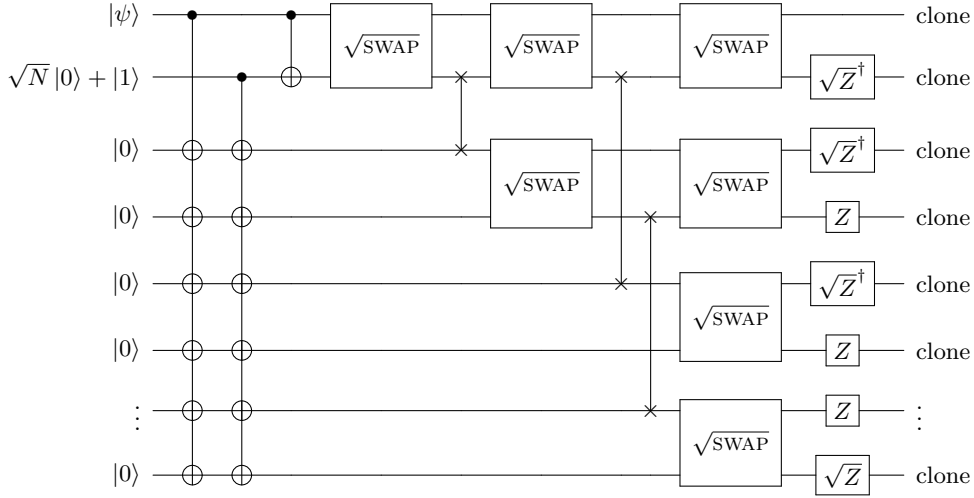

All of the complexity of producing these clones is conveniently wrapped up in just two helper functions. The corresponding circuit diagrams are contrasted in Figs.\ \ref{fig:QcircuitHam} and \ref{fig:QcircuitGates}. While a quantum circuit for cloning has previously been explicitly stated for small sizes \cite{buzek1997}, we are not aware of a version, other than probabilistic versions \cite{kay2016a}, that works deterministically for general $1\rightarrow N$ universal symmetric cloning, let alone the asymmetric case. This is probably because the cloning map in Eq.\ (\ref{eqn:clonemap}), specialised to symmetric cloning, is not the map usually stated \cite{gisin1997,murao1999}: \cite{kay2009a} reveals that the cloning map is associated with the ground state of a particular matrix, and the symmetric case is highly degenerate. The version that we have chosen, Eq.\ (\ref{eqn:clonemap}), extends consistently from the asymmetric case, and lends itself well to implementation with a quantum circuit, as depicted in Fig.\ \ref{fig:QcircuitGates} provides such a definition for symmetric cloning. This circuit can be modified for asymmetric cloning. Assuming that the architecture exhibits only nearest-neighbour couplings (the only instance where it makes sense to consider implementation via a nearest-neighbour Hamiltonian), the depth of the circuit is $N$, and comprises $O(N^2)$ gates.

Why do we only create clones on every second site of the chain? Imagine we have a Hamiltonian like Eq.\ (\ref{eqn:exchange}), but including magnetic fields as well.
\begin{equation}
H=\sum_{n=1}^{M-1}\frac{J_n}{2}(X_nX_{n+1}+Y_nY_{n+1})+\sum_{n=1}^N\frac{B_n}{2}(\identity-Z_n)	\label{eqn:exchange2}
\end{equation}
Let
$$
U=\prod_{n=1}^{(N+1)/2}X_{2n-1}\prod_{n=1}^{(N-1)/2}Y_{2n}.
$$
We have that $UHU=-H+\left(\sum_{n=1}^NB_n\right)\identity$. Moreover, at $t_0$, $e^{-iHt_0}=e^{iHt_0}$ because $e^{-i\lambda_nt_0}=\pm 1$ for every eigenvalue (neglecting, for simplicity, a possible global phase). Thus, time evolution in the higher excitation subspace is given by
$$
e^{-iHt_0}\ket{\bar n}=-(-i)^{(N-1)/2+2n}Ue^{-i\left(H+\left(\sum_{n=1}^NB_n\right)\identity\right)t_0}\ket{n}.
$$
If the state produced in the single excitation subspace is
$$
\ket{\psi}=\sum_n\alpha_n\ket{n},
$$
in the higher excitation subspace we get
\begin{multline*}
(-i)^{(N+3)/2+2n}e^{-it_0\sum_mB_m}U\sum_m\alpha_m\ket{m}\\=e^{-it_0\sum_mB_m}\sum_m(-1)^{n+m}\alpha_m\ket{\overline m}.
\end{multline*}
This can only be consistent with the desired transform if $\alpha_m=0$ on every second site, to eliminate the effect of the $(-1)^{n+m}$ term. In doing so, it transpires that for a symmetric target spectrum, it suffices to set $B_n=0$. One can readily verify that although the even-numbered qubits effectively act as ancillas, and although the transformation we implement does not leave them separable, it does not adversely affect the cloning fidelity. Another way to circumvent this restriction is to introduce more parameters to the Hamiltonian. For instance,
$$
H_{XXZ}=\sum_{n=1}^{M-1}\frac{J_n}{2}(X_nX_{n+1}+Y_nY_{n+1})+B_nZ_nZ_{n+1}
$$
commutes with $X^{\otimes N}$, and gives us the ability to manipulate the diagonal elements in the single excitation subspace via the parameters $\{B_n\}$. This gives us sufficient control to produce a system that gives clones on every site of the chain.

\subsection{Designing State Synthesis Systems} \label{sec:synthesis}

While we can use the algorithms of \cite{kay2017,kay2017a} to generate figures such as Fig.\ \ref{fig:WstateOddRevival}, these are limited by very small radii of convergence. Instead, we would now like to examine if the isospectral flow ideas outlined above can be applied. The two papers \cite{kay2017,kay2017a} provide different philosophies for how to produce a useful chain, but for our purposes, much of the calculation in the same.

Let us start with a first guess at a Hamiltonian, $H_0$ (in the single excitation subspace). It has a desired spectrum but its evolution produces
$$
e^{-iH_0t_0}\ket{\phi}=\ket{\psi_0},
$$
starting from the separable state $\ket{\phi}$ (i.e.\ a single excitation on a particular site), and evolving for a time $t_0$, where $\ket{\psi_0}$ is not our target state $\ket{\psi_t}$. In practice, we will set $\ket{\phi}=\ket{\frac{N+1}{2}}$ to minimse the evolution time. However, if $\ket{\psi_t}$ is symmetric, we can reduce the task to finding a chain of half the length, starting with an excitation at site 1 \cite{kay2017a}. For that reason, we will typically assume $\ket{\phi}=\ket{1}$.

How can we make a better guess, $H_1$, which should have the same spectrum as $H_0$? Again, we use the isospectral transformation
$$
H_1=e^{-B}H_0e^{A}
$$
for some anti-Hermitian matrices $A$, $B$. We have several properties that we want to impose. As in the Ising case, we can rearrange the matrix $H_0$ into a block structure of \{all odd elements, all even elements\}, so that $H_0$ takes the form
$$
H_0=\left(\begin{array}{cc}
0 & X_0 \\ X_0^T & 0
\end{array}\right).
$$
A block-diagonal $A=\text{diag}(A_o\ A_e)$ preserves the structure of $H_0$, with $X_0$ evolving as $X_0\rightarrow e^{-A_e}X_0e^{A_o}$.

Next, we want to impose that the tridiagonal structure of $H_0$ is preserved. This just requires $\bra{i}\dot H_0\ket{j}=0$ for all $|i-j|>1$. We've already partially achieved this with our block-diagonal of $A$ ensuring that it's true for all $|i-j|$ even. The remainder are simply a set of simultaneous linear equations in the parameters $A_o$ and $A_e$:
$$
\bra{i}X_0A_o-A_eX_0\ket{j}=0 \forall j\neq i,i-1.
$$ 

We are hence building up a set of linear conditions which, so far, just ensure that subsequent matrices maintain the important properties of the initial matrices. Now we must impose that each subsequent iteration moves us towards a better evolution. Since we are updating
$$
H_0\mapsto e^{-A}H_0e^A,
$$
and assuming $A$ is small, the evolution updates as
$$
e^{-A}e^{-iH_0t_0}e^{A}\ket{\phi}\approx \ket{\psi_0}-A\ket{\psi_0}+e^{-iH_0t_0}A\ket{\phi}.
$$
To fix the the new evolution to be $\ket{\psi_t}$, we might solve
$$
-A\ket{\psi_0}+e^{-iH_0t_0}A\ket{1}=\ket{\psi_t}-\ket{\psi_0}
$$
subject to the structure constraints that we have already described. However, a solution of this form is unlikely to keep $A$ is small, it is perhaps better to describe it as a constrained optimisation problem (linear programming):
$$
\max_{A:\|A\|\leq \delta}(\bra{\psi_t}-\bra{\psi_0})[e^{-iH_0t_0},A]\ket{\phi}.
$$
Having found $A$, we update $H_0$ according to the unitary transformation update (rather than direct integration).

\begin{figure}[!tbp]
\begin{center}
\includegraphics[width=0.4\textwidth]{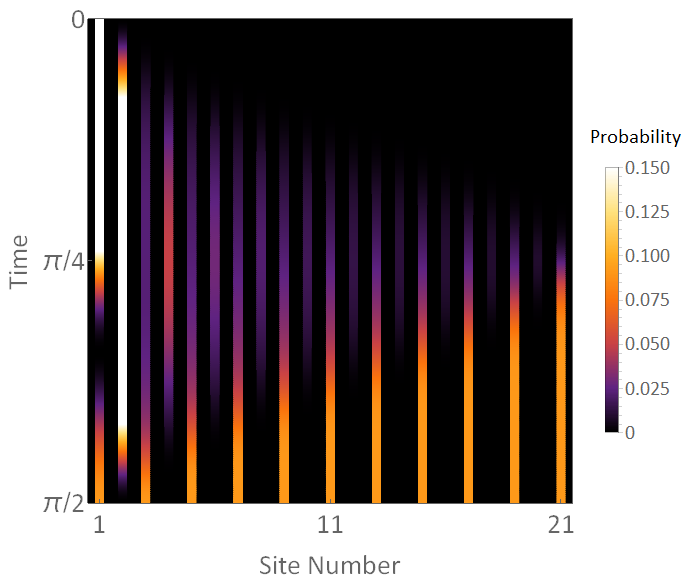}
\end{center}
\vspace{-0.5cm}
\caption{A single excitation, input to the first spin of a 21-qubit spin chain (top) can be caused to evolve into a uniform superposition over the odd-numbered sites (bottom), as required for symmetric $1\rightarrow 11$ cloning.}\label{fig:WstateOdd}
\vspace{-0.3cm}
\end{figure}

For the particular design philosophy of \cite{kay2017}, we can go further. There, due the the chosen spectrum,
$$
e^{-iH_0t_0}=\identity-2\proj{\lambda_0},
$$
the aim is to fix the null vector to $\ket{\lambda_t}$, where $$\ket{\psi_t}=(\identity-2\proj{\lambda_t})\ket{1},$$ and knowing that
the null vector of the next iteration is $e^{-A}\ket{\lambda_0}$. In successive iterations, we aim to maximise the overlap with $\ket{\lambda_t}$. %If we bound that every element of $A$ is less than some parameter $\delta$ ($\|A\|_\infty\leq\delta$), then for sufficiently small $\delta$, we expand the overlap to first order, and thus effectively 
Hence, we have to solve the linear programming problem
$$
\min_{\|A\|_\infty\leq\delta}\bra{\lambda_t}A\ket{\lambda_0}.
$$
Convergence is well motivated -- unless there is a reason that either $\ket{\lambda_0}$ or $\ket{\lambda_t}$ must be an null vector of $A$, we can always find a non-zero value of the overlap, and choosing the sign of $A$ assures that the outcome is always negative, and hence iterates towards an improved solution. The solution must converge, and further justification that it converges globally on the correct solution is given in Appendix \ref{sec:appendix}. A typical output is shown in Fig.\ \ref{fig:WstateOdd}, and is used in \cite{kay2017b} to demonstrate the proper functioning of the entire evolution sequence, giving the optimal symmetric cloning fidelity of $F=\frac{23}{33}$.

\subsubsection{Hypercubes} \label{sec:hypercube}

So far, we have discussed engineering a one dimensional chain, choosing the coupling strengths to achieve the evolution that we desire. This perfectly parallels studies of perfect state transfer. However, another avenue for study in perfect state transfer is the set of uniformly coupled graphs that can accomplish the same task. For example, hypercubes of arbitrary dimension $k$, and side length 2 or 3, achieve perfect state transfer at distances $k$ or $2k$ respectively \cite{christandl2005}. Various other graphs have since been shown to provide perfect transfer, including a variety of modifications of the hypercubes \cite{christandl2005,angeles-canul2010,bernasconi2008,facer2008}. Can graphs also be used for the state synthesis tasks that could be useful for quantum cloning? We specifically focus on generating uniform superpositions across some subset of sites, with a preference for those where the phase on each of the superposed sites is the same. 

Let $G$ be a graph with edges $E$ and $N$ vertices $V$. The graph Hamiltonian is defined as
$$
H_G=\half\sum_{(i,j)\in E}(X_iX_j+Y_iY_j).
$$
As before, there is a subspace structure based on the number of excitations, and $\ket{0}^{\otimes N}$ and $\ket{1}^{\otimes N}$ are null vectors, and $[H,X^{\otimes N}]=0$ -- we only have to get the evolution in the single excitation subspace correct. If the graph is bipartite, the phase choice can only be consistent if the superposed sites are all part of the same bipartition, by a  generalisation of the argument around Eq.\ (\ref{eqn:exchange2}).

In the single excitation subspace, the Hamiltonian is represented by the adjacency matrix, $A$, of the graph. The conditions on state synthesis \cite{kay2017a} are remarkably similar to those of perfect transfer \cite{kay2010a,kay2011a}: to start from a site $n$, producing a state $\ket{\psi}$ in time $t_0$, if $\ket{\psi}\in\mathbb{R}^{N}$, then
$$
\braket{\lambda_m}{n}=\pm\braket{\lambda_m}{\psi}
$$
for every eigenvector $\ket{\lambda_m}$, and for those eigenvectors for which $\braket{\lambda_m}{n}\neq 0$, the eigenvalues $\lambda_m$ must satisfy
$$
e^{-i\lambda_mt_0}=\pm e^{i\phi}
$$
for some phase $\phi$. This has some further consequences for the spectrum of the adjacency matrix $A$ \cite{godsil2012a}. For example, with one extra assumption about the nature of the state synthesis task (that all vertices have a perfect revival at the same time), we know that the spectrum for a non-bipartite graph must be integral, while for a bipartite graph, the spectrum is either integral, or rational multiples of $\sqrt{\Delta}$ for a square-free integer $\Delta$. We will not develop this theory more generally here, but will focus on some special cases that we have found.

Several instances of the path $P_n$ (i.e.\ uniformly coupled chain of $n$ vertices) generate uniform superpositions:
\begin{eqnarray*}
e^{-iA(P_2)\pi/4}\ket{1}&=&\frac{\ket{1}-i\ket{2}}{\sqrt{2}}	\\
e^{-iA(P_3)\pi/\sqrt{8}}\ket{2}&=&\frac{\ket{1}+\ket{3}}{\sqrt{2}}	\\
e^{-iA(P_3)\cos^{-1}\left(\frac{1}{\sqrt{3}}\right)/(\sqrt{2}\pi)}\ket{2}&=&\frac{\ket{1}+i\ket{2}+\ket{3}}{\sqrt{3}}	\\
e^{-i2A(P_5)/\sqrt{27}}\ket{3}&=&\frac{\ket{1}+i\ket{2}+i\ket{4}+\ket{5}}{2}
\end{eqnarray*}
The second of these is ideally suited to $1\rightarrow 2$ symmetric cloning (and is closely related to a previous construction \cite{dechiara2004,chen2006}). We can extend these cases by using the hypercube construction \cite{christandl2005}. For a graph $G$, the adjacency matrix of the corresponding $k$-dimensional hypercube is
$$
A(G^k)=\sum_{n=1}^k\identity_N^{\otimes(n-1)}\otimes A\otimes\identity_N^{\otimes (k-n)}.
$$
This describes independent evolution along each of the $k$ dimensions. Taking a basis $x\in[N]^k$ (a $k$-dimensional vector where each element takes an integer value from $1$ to $N$), each $\ket{x}$ corresponds to a single excitation being on a particular vertex of the graph. Starting from $\ket{y}$, the final amplitude on a vertex $\ket{x}$ is
$$
\prod_{i=1}^k\bra{x_i}e^{-i At}\ket{y_i}.
$$
If $A$ gives a uniform superposition, so does the hypercube. The hypercube of side length 2 (i.e.\ derived from $P_3$) is particularly compelling. For example, a $3\times 3$ square lattice of uniformly coupled qubits generates a uniform superposition of all 9 sites. Or, more applicable to symmetric cloning, $P_3^k$ produces (at a different time) the uniform superposition across all $2^k$ corners.

\section{Conclusions}\label{sec:conclude}

We have shown how a fixed transverse Ising system can produce a GHZ state, which is a key quantum resource for use in future technologies. This could sit as a stand-alone device, or as a special unit, a co-processor, within a larger quantum device. The fixed-function co-processor replaces what would otherwise be a complex sequence of unitary gates, with the inaccuracies inherent in the multiple separate steps that have to be taken in its implementation. By extending the results from $H_I$ to $H_{ZY}$, we have potentially opened GHZ synthesis to a much broader range of experiments. Realising that the transformation required to achieve GHZ state synthesis in a transverse Ising model essentially reduces to a state-transfer-like condition on $h_1$ would also significantly simplify optimal control studies such as \cite{wang2010a}, moving away from the assumption of perfect engineering.
%More generally, if the perfect transfer coupling scheme is replaced with one that realises a perfect revival at either end of the chain \cite{kay2010a}, i.e.\ for any real value $\gamma$
%$$
%e^{-i\tilde h_1t_0}\ket{n}=\cos\gamma\ket{n}+(-1)^n\sin\gamma\ket{2N+1-n},
%$$
%then in the transverse Ising model, an evolution for time $t_0/2$ implements the transformation 
%$$
%\ket{0}^{\otimes N}\mapsto\cos\frac{\gamma}{2}\ket{0}^{\otimes N}-i\sin\frac{\gamma}{2}\ket{1}^{\otimes N}.
%$$

We have also specified a second transformation. This two step procedure, where the first step uses the GHZ synthesiser, implements optimal asymmetric universal cloning of qubits. This is the first time that a non-probabilistic strategy has been given for these cloning machines. Our transformation is implemented by operations that are local in a one-dimensional chain of qubits. Our use of the GHZ synthesiser to implement a single-control, multiple target controlled-not gate may be of further interest in the relation to the Fourier transform.

Both transformations, when restricted to a nearest-neighbour architecture, exhibit an optimal $O(N)$ scaling in run-time, and have essentially identical running times to their quantum circuit equivalents.

Central to these results was a new isospectral transformation algorithm, with fine-grained control over directing consecutive iterations. Global convergence of the algorithm is well-motivated. Mathematica scripts that implement the reported results for chains of 21 qubits are available from \cite{kay2017b}. The algorithm demonstrates considerable potential for further development, and should be broadly applicable.

{\em Acknowledgements}: We would like to thank L.\ Banchi and G.\ Coutinho for introductory conversations. This work was supported by EPSRC grant EP/N035097/1.

%\bibliography{../../../References}	%Transfer the references from the bib file before submission of electronic version (this is probably upon acceptance)
%merlin.mbs apsrev4-1.bst 2010-07-25 4.21a (PWD, AO, DPC) hacked
%Control: key (0)
%Control: author (8) initials jnrlst
%Control: editor formatted (1) identically to author
%Control: production of article title (-1) disabled
%Control: page (0) single
%Control: year (1) truncated
%Control: production of eprint (0) enabled
%

\appendix
\section{Convergence of Isospectral Algorithm}\label{sec:appendix}

%\subsection{Convergence} \label{sec:converge}

We aim to motivate that the algorithm described in Sec.\ \ref{sec:synthesis} has a single point of convergence provided the target null vector, $\ket{\lambda_t}$, can be a null vector of a system with the fixed spectrum.

At each step, we iterate with a matrix $A$, and impose that the elements of $A$ are bounded by some step size $\delta$, $\|A\|_\infty\leq\delta$. It is important that we pick $\delta$ such that the second order term in the expansion of $\bra{\lambda_t}e^{-A}\ket{\lambda_0}$ is negligible. Since $\bra{\psi}A\ket{\psi}=0$ for all real-valued states $\ket{\psi}$, the maximum value of $\bra{\lambda_t}A\ket{\lambda_0}$ is achieved with 
$$A\ket{\lambda_0}\sim\ket{\lambda_t}-\braket{\lambda_0}{\lambda_t}\ket{\lambda_0}.$$
This is generically possible to fix: $A\ket{\lambda_0}$ has $(N+1)/2$ components, and the $(N-1)/2$ free parameters can control the output in the space orthogonal to $\ket{\lambda_0}$. By aligning these vectors with the left- and right- maximum singular vectors of $A$ (singular value $\sigma$), we have $\bra{\lambda_t}A\ket{\lambda_0}=-\sigma\sqrt{1-\braket{\lambda_0}{\lambda_t}^2}$. Meanwhile, $\bra{\lambda_t}A^2\ket{\lambda_0}<\sigma^2$, i.e.\ $\sigma\ll\sqrt{1-\braket{\lambda_0}{\lambda_t}^2}$. Through the following bounds, we relate $\delta$ to $\sigma$:
$$
\sqrt{\frac{N+1}{2}}\delta\geq\sqrt{\frac{N+1}{2}}\|A\|_\infty\geq \|A\|_2\geq\sigma,
$$
where $\|A\|_2$ is the Frobenius norm. Thus, by ensuring that $\delta=\epsilon\sqrt{1-\braket{\lambda_0}{\lambda_t}^2}$ for small $\epsilon>0$, the second order term is always negligible. Indeed, we can directly bound the value $\chi=\braket{\lambda_t}{\lambda_0}$:
$$
\frac{d\chi}{dt}\leq -\sigma\sqrt{1-\chi^2}\leq-\epsilon(1-\chi^2).
$$
This solution tends exponentially towards $\chi=1$. Assuming that $A\ket{\lambda_0}$ can be picked as specified, we have convergence, and accuracy $\varepsilon$ is achieved with an average complexity of $O(N^6\log(\sqrt{N}/\varepsilon))$, the leading term arising from solving $O(N^2)$ linear constraints.

Generically, $A\ket{\lambda_0}$ can be any state in the odd space that is orthogonal to $\ket{\lambda_0}$, since there are $(N-1)/2$ free parameters with which to achieve this. So, when does this fail? Is this compatible with the observation that some states cannot be the null vector for a tridiagonal system of a particular spectrum \cite{kay2017}?

We start by noting that although \cite{kay2017} indicated that none of the vector elements on the odd space can be 0 (as this would give consecutive 0 elements on the complete vector), this was an artificial imposition resulting from requiring non-zero coupling strengths. However, this consideration is not built into the algorithm, so we are not prevented from reaching these forms of $\ket{\lambda_t}$.

Since the state $A\ket{\lambda_0}$ is linear in the free parameters, the space described when the corresponding vectors do not span the space (and are hence linearly dependent on each other) must be a convex space. There is a single inaccessible region, which must therefore include any inaccessible null vectors. The only question is whether this region is tight with that of the inaccessible null vectors.

\subsection{Case Study} \label{sec:case}

In the absence of universal answers, we investigate the special case of $N=5$ and spectrum $0,\pm3,\pm5$, since this is a case where there are forbidden null vectors \cite{kay2017}. The odd space is dimension 3, and there are two free parameters $a$ and $b$. For a particular $h$ with coupling strengths $J_1,\ldots,J_4$,
$$
A_o\ket{\lambda_0}=a\left(\begin{array}{c}
J_1J_2(J_3^2+J_4^2) \\ J_1^2J_3^2-J_2^2J_4^2 \\J_3J_4(J_1^2+J_2^2)
\end{array}\right)+b\left(\begin{array}{c}
J_1(J_3^2+J_4^2-J_1^2) \\ -J_2(J_1^2-J_4^2) \\ -J_2J_3J_4
\end{array}\right).
$$
The ratios $\gamma_1=\frac{J_1}{J_2}$ and $\gamma_2=\frac{J_4}{J_3}$ parametrise the possible null vectors
$$
\ket{\lambda_0}=\left(\begin{array}{c}
J_2J_4 \\ -J_1J_4 \\ J_1J_3
\end{array}\right),
$$
from which we can derive that the only time that we do not have access to the whole space is when $J_1^2+J_2^2=J_3^2+J_4^2$, and thus
\begin{equation}(1+\gamma_1^2)(1+\gamma_2^2)=\frac{17^2}{8^2},\label{eqn:boundary}\end{equation}
having used the eigenvalue relations (such as $J_1^2+J_2^2+J_3^2+J_4^2=34$) to eliminate the remaining terms. This defines a barrier in the possible space of $\ket{\lambda_t}$ that the algorithm cannot cross.

Now let us consider the region of $\ket{\lambda_t}$ for which there is no $h$ with the correct spectrum and that null vector. Using the explicit eigenvalue relations for the coupling strengths, we can write that
$$
(J_2^2(1+\gamma_1^2)-17)^2=\left(64-\frac{J_2^2}{1+\gamma_2^2}(34-J_2^2(1+\gamma_1^2))\right).
$$
This has a non-negative solution for $J_2^2$ when 
$$(1+\gamma_1^2)(1+\gamma_2^2)\geq\frac{17^2}{8^2}.$$
We conclude that our algorithm is capable of converging on any valid null vector for $N=5$. If we demand an invalid null vector, the algorithm will converge somewhere on the surface of closest approach, defined by Eq.\ (\ref{eqn:boundary}). It is reasonable to expect similar behaviour in larger spaces, but this remains unproven.

\end{document}